\documentclass[preprint,12pt]{aastex}

\shorttitle{Envelope of L1527}
\shortauthors{Tobin et al.}
\begin{document}

\title{The Inner Envelope and Disk of L1527 Revealed: Gemini L$^{\prime}$-band Scattered Light Imaging}
\author{John J. Tobin\altaffilmark{1},  Lee Hartmann\altaffilmark{1}, \& Laurent Loinard\altaffilmark{2}}
\altaffiltext{1}{Department of Astronomy, University of Michigan, Ann
Arbor, MI 48109; jjtobin@umich.edu}
\altaffiltext{2}{Centro de Radioastronom{\'\i}a y Astrof{\'\i}sica, UNAM, Apartado Postal
3-72 (Xangari), 58089 Morelia, Michoac\'an, M\'exico}

\begin{abstract}

We present high-resolution L$^{\prime}$-band imaging of the 
inner scattered light structure of Class 0 protostar 
L1527 IRS (IRAS 04368+2557) taken with the Gemini North telescope.
The central point-source like feature seen in \textit{Spitzer Space Telescope} 
IRAC images is resolved in the Gemini image
into a compact bipolar structure with a narrow dark lane in the center.
Two scattered light lobes are extended
$\sim$1.8$^{\prime\prime}$ (200 AU) perpendicular to the direction of
the outflow and $\sim$2.5$^{\prime\prime}$ (350 AU) 
along the outflow axis; the narrow dark lane between the
scattered light lobes is $\sim$ 0.45$^{\prime\prime}$ (60 AU) thick. 
The observations are consistent with our initial modeling 
of a bright inner cavity separated by a dark lane due to extinction
along the line of sight of the central protostar by the disk (Tobin et al. 2008).
The bright, compact scattered light might be due to complex inner structure
generated by the outflow, as suggested in our first paper, or it
may more likely be the upper layers of the disk forming from infalling matter.

\end{abstract}


\section{Introduction}

In the earliest stages of star formation, the Class 0/I phases \citep{andre1993,lada1987},
the newborn protostar is embedded within a dense envelope of gas and dust. 
In the Class 0 phase, little or no emission is generally detected shortward of $\sim$10$\mu$m
due to high extinction toward the central object; however, 
the bipolar outflows from the central protostar and disk
carve cavities in the envelopes creating scattered light nebulae visible in the near to mid-infrared (1-8$\mu$m).
The morphology of the outflow cavities then result from a ram pressure
balance between the infalling envelope
and outflow \citep{shu1991}.

Early models of infalling envelopes assumed spherical collapse \citep{shu1977} along with
slow rotation \citep{ulrich1976,cassen1981,tsc1984}, which were later
modified for infall from initially flattened geometries \citep{hartmann1996}. 
The rotating collapse model provides a natural route for the formation
of circumstellar disks within the protostellar envelope. However, disks in Class 0 objects
have been difficult to study primarily because of the surrounding envelope \citep[e.g.][]{jorgensen2009} 
and insufficient resolution at millimeter wavelengths to isolate the protostellar disk.
In addition, the small-scale structure of 
outflow cavities in Class 0 protostars has also been elusive due to the envelope extincting near-IR
scattered light emission, shifting the scattered light emission peak into the mid-IR \citep{tobin2007,tobin2008}.

One of the nearest low-mass protostars in Taurus (d = 140pc), L1527 IRS (hereafter L1527), 
has been a favorite target for studies of outflow and envelope structure \citep[e.g.][]{ohashi1997,
chandler2000,rob2007,tobin2008,gramajo2010}. It has been classified as Class 0 protostar \citep{chen1995}
but could be a Class I system given its edge-on orientation \citep{ohashi1997}, which
enhances the amount of extinction along the line of sight \citep{tobin2008}.
The moderate resolution IRAC images of L1527
from \textit{Spitzer} reveal bright bipolar cavities extending $\sim$10000 AU in radius \citep{hartmann2005,tobin2008}.
The outflow cavities are separated by what appears to be a large-scale ($\sim$1000 AU) dark lane; however,
we observed a central point-source between the cavities within the dark lane from 3.6 to 8$\mu$m.
In \citet[][hereafter Paper I]{tobin2008}, we were able to explain this by constructing
a model assuming that the point-source
was a compact inner outflow cavity, unresolved by \textit{Spitzer}, connecting with the
larger outer cavity. We then proposed high-resolution mid-IR imaging to test
this model.

In this paper, we present high resolution L$^{\prime}$-band (3.8$\mu$m) observations of L1527  
from the Gemini North telescope. This is the highest resolution mid-IR image 
of the inner envelope/outflow cavity around a Class 0 protostar. The observations
confirm our prediction of a compact, bipolar scattered light structure; quite similar to the 
model constructed in Paper I but strongly resembling a disk in scattered light.

\section{Observations and Data Reduction}

We observed L1527 with the 8.1 meter Gemini North telescope at Mauna Kea on 2009 October 24 and 2009 December 11
using the Near-Infrared Imager (NIRI) in L$^{\prime}$-band (3.8$\mu$m). The camera was used in f/14 mode which
provides a 0.049$^{\prime\prime}$ pix$^{-1}$ scale; we only used the central 512x512 area
of the detector for faster read-out which enabled integrations
short enough to avoid saturation from sky background. We observed with a standard 5-point ``dice-5''
dither pattern with 5$^{\prime\prime}$ steps and a nearby star was used for tip-tilt correction and guiding.
At each position 150 coadded images of 0.2 seconds were taken; a total of 35 images were
taken yielding 16 minutes of integration during the first observation
and 176 images totaling 88 minutes of integration during the second observation. However, the 
first 38 images in the second observation were unusable due to variable sky background. Observations
of the standard star HD 22686 show that the seeing was $\sim$0.3$^{\prime\prime}$ on both nights.

The raw data were reduced using the Gemini IRAF\footnote{IRAF is distributed by the National
Optical Astronomy Observatories,
which are operated by the Association of Universities for Research
in Astronomy, Inc., under cooperative agreement with the National
Science Foundation.} package. Flat field images were generated from sky flats
constructed by median combining the dithered images. Each on-source
frame was sky subtracted using a median sky generated from 
the on-source frame and the two images taken immediately before and after. This method of 
sky subtraction compensates for the
rapid variations in sky brightness at L$^{\prime}$-band. The sky subtracted images were then combined using the imcombine
task of IRAF using the offsets given in the header; L1527 was quite faint in the individual frames and could not be used
for further refinement of the offsets. The datasets from both epochs of observation were combined
yielding a final image with 85 minutes of integration. 

We also include data taken with the Infrared Array Camera (IRAC) \citep{fazio2004} on the \textit{Spitzer Space Telescope}. These data 
were taken in 24 February 2004 as part of the GTO Taurus survey and  were presented in \citet{tobin2008} and \citet{hartmann2005}.

\section{Results}

The images of L1527 at L$^{\prime}$-band and 3.6$\mu$m are shown in Figure 1.
 The IRAC image clearly shows the point-like structure appearing in the center
 of the envelope between the large scattered light cavities. The Gemini L$^{\prime}$ 
image resolves the point-like structure into a compact, bipolar scattered
light nebula at the center of L1527. The two lobes are separated by a 
narrow dark lane $\sim$0.45$^{\prime\prime}$ (60 AU) wide, consistent with a circumstellar
disk shadow. The L$^{\prime}$ image reveals structure similar to that of the 
\textit{Hubble} NICMOS images of Class I protostars in Taurus from \citet{padgett1999}. 
However, this is the first time such an image has been captured
of a Class 0 protostar. 

The total extent of the bipolar structure in the direction of the outflow (east-west)
is $\sim$2.5$^{\prime\prime}$ (350 AU) and $\sim$1.8$^{\prime\prime}$ (250 AU) in width.
The extent of the scattered light along the outflow and dark lane is much thicker than a prototypical disk
\citep[i.e. HH30,][]{burrows1996} indicating that the disk in L1527 is more vertically extended or
that the scattering is coming from the inner envelope.
The eastern lobe is about twice as bright as the western lobe;
consistent with eastern side being slightly inclined toward us ($\sim$85$^{\circ}$) (Paper I). The orientation of the disk
dark lane indicates that the rotation axis of the system 
has a position angle $\sim$85$^{\circ}$ east of north. We also see extended,
low surface brightness outflow cavity emission extending away from the
inner envelope in the Gemini image. The cavity is quite narrow until
$\sim$6$^{\prime\prime}$ (840 AU) from the center; this narrowness
was not well resolved in the \textit{Spitzer} image. Past 6$^{\prime\prime}$,
the cavity expands rapidly to very wide angles.
The total flux within a 7$\farcs$14 (1000 AU) aperture radius
is 6.5 $\pm$ 0.6 mJy, consistent with the IRAC flux from Paper I.

\subsection{Models}

We use the Monte Carlo radiative transfer code of \citet{whitney2003} to interpret the observations.
The model components are described in Paper I, but we discuss some important aspects here 
for completeness. The envelope density structure is the standard rotating collapse
solution \citep{cassen1981,tsc1984} in which envelope density scales with
the infall rate $\dot{M}_{env}$ with $\rho$ $\propto$ R$^{-3/2}$ outside
the centrifugal radius (R$_C$) and $\rho$ $\propto$ R$^{-1/2}$ inside R$_C$. 
Conical outflow cavities are carved out of the envelope, their shape is defined by a polynomial,
the degree of the polynomial is the shape parameter. We use the dust model from Paper I 
for the envelope which has grains up to 1$\mu$m in radius.

The disk is defined by its initial scale height H$_{0}$ at the stellar radius R$_{*}$ and
flaring power law H $\propto$ R$^{\beta}$, radial density profile $\rho$ $\propto$ R$^{-\alpha}$,
mass, and outer radius. The vertical structure of the disk is assumed to
be Gaussian with an initial scale height defined to be in hydrostatic
equilibrium at the dust destruction radius (T$_d$=1600K).
Values of $\beta$ = 9/8, 5/4 are commonly used, corresponding to
isothermal scale heights with $T_{disk}$ $\propto$ R$^{-3/4}$
and R$^{-1/2}$. 
The gas and dust of the disk
are assumed to be well mixed. The dust opacities for the disk are 
taken from \citet{wood2002} which were used to model the SED of HH30. The dust is distributed in a 
quasi-settled manner with grains up to 1mm in the disk midplane while smaller
grains remain extended. The transition between these dust populations
is defined to be n$_{H_2}$=10$^{10}$cm$^{-3}$ or the first scale height.


In Paper I, we inferred that the apparent point source seen at 3.6$\mu$m was in fact a bipolar
structure seen in reflected light, with the true central source actually hidden by extinction.
To create this structure, we constructed a dual-cavity model having a narrow inner cavity with an outer
cavity offset by 100 AU (see Figure 10 of Paper I for an illustration).
Comparing the top and middle rows of Figure 2
shows that our prediction was qualitatively correct.  We next take advantage of the
high spatial resolution of the Gemini data to improve our understanding of 
the inner structure.  We required that models reproduce both the L$^\prime$ and IRAC images 
and the broadband SED and IRS spectrum \citep[][and Paper I]{furlan2008}, as
it is important to use as many constraints as possible \citep[Paper I,][]{gramajo2010}.

We first adjusted the parameters of the dual-cavity model 
to better reproduce the Gemini observations by
varying the inner and outer cavity shapes and opening angles.  As shown in
Figures 2 and 3, our best-fit dual-cavity model reproduces the observations reasonably
well (note that the images for the refined dual-cavity model are not shown as they
nearly identical to the disk model as discussed below). 
The envelope infall rate for this model
was 10$^{-5}$M$_{\sun}$ yr$^{-1}$; the outer cavity was offset from the inner cavity
by 85 AU. The disk parameters are H(10AU)= 1.87 AU with $\beta$ = 1.25,  H$_{0}$ = 0.0332 R$_{*}$,
$M_{disk}$= 0.05M$_{\sun}$ and $R_{disk}$= 25 AU; a model with $\beta$ = 1.125
was also able to be fit, but with double the initial scale height; the full parameter set is listed
in Table 1.

We did not consider a disk model in Paper I because the large-scale dark lane seemed to
be too thick to be reproduced with a disk. However, using the detailed Gemini image, we can describe disk parameters
which yield scattering surfaces comparable to that of the dual-cavity model.
The highly-flared disk in this model had H(10AU)= 1.95 AU (H(190AU) = 82 AU),
with $\beta$ = 1.27, $\alpha$ = 3.0, H$_{0}$ = 0.03 R$_{*}$, $M_{disk}$= 0.005M$_{\sun}$
and $R_{disk}$= 190 AU. The 
envelope infall rate was lowered to 0.8 $\times$ 10$^{-5}$M$_{\sun}$ yr$^{-1}$ in
order to allow more scattered light from the disk to escape through the envelope.
We also used the envelope dust model for the disk upper layers rather than the default dust model;
this increased the L$^{\prime}$ intensity
and reduced emission from 10 to 60 $\mu$m (see Table 1 for other parameters) bringing the model
into closer agreement with the observations. 
As shown in Figures 2 and 3, this model also provides reasonable agreement with the observations.

\section{Discussion}

The dual-cavity model in Paper I was justified
based on simulations with a wide-angle outflow and a 
highly flattened envelope density distribution 
\citep{delamarter2000}.  However, it is unclear if such
a flattened envelope is present in L1527, as there is no evidence for this in
submillimeter observations \citep{chandler2000}.
In addition, the outflow cavity on scales larger that 840 AU from the central object
does not easily fit within this framework due to the rapid
widening of the scattered light nebula beyond 840 AU (\S 3; Figure 1) 
This is difficult to explain with a simple conical cavity
because there is no direct line-of-sight from the protostar/disk to these parts of
the outflow cavity.  Therefore we suggest that the morphology of the scattered light
nebulosity on large scales is not due directly to the outflow but to the
morphology of the envelope and/or ambient medium \citep{tobin2010}.
In any case it appears difficult to explain the complex shape of the 
scattered light structures with an simple outflow-envelope interaction.

Given these difficulties, the disk scenario seems attractive.
Images of disks in silhouette have previously been observed around T-Tauri stars (Class II sources) and Class I protostars
\citep[e.g.][]{padgett1999,burrows1996}. The small-scale scattered light morphology of 
L1527 bears a resemblance to that of IRAS 04302+2247 and HH30. 
The problem for this model is that the disk must be highly flared, with the aspect ratio
of the disk extinction lane relative to the extent of the scattered light surfaces being about a factor
of two larger than other objects. To reproduce the compact scattered light and large-scale 
dark lane with a disk, it was necessary to adopt a high degree of
flaring in the disk ($\beta$=1.27), a steep density profile
($\alpha$=3.0), a large disk radius (R$_d$=190AU), and a small disk mass (M$_d$=0.005M$_{\sun}$).
(Note that the mass is simply a formal parameter of the model for scaling the density.
Extrapolating disk masses from fitting scattered light in the disk upper
layers is problematic since they contain a small fraction of the total mass.)
These observations and modeling emphasize not only that disks form 
early during protostellar collapse \citep{jorgensen2009,vorobyov2009} but 
also that disks with large radii are able to form during the Class 0 phase
\citep[see also][]{enoch2009} with radii comparable
to Class II disks in Taurus \citep{andrews2007}.

In general, dust growth and/or depletion in upper disk
layers relative to ISM conditions is needed to reproduce the scattered light images
of edge-on disk systems and the SEDs of T Tauri stars \citep{dalessio2006,duchene2010,furlan2008}. 
Thus, it may be that the postulated L1527 disk has had much less dust evolution and
thus has more small dust in upper layers, making it easier to explain the
wide extinction lane observed \citep[e.g.][]{dalessio2006}.

\citet{loinard2002} directly imaged the disk of L1527 and a 
binary companion separated from the primary by 0.17$^{\prime\prime}$ (24 AU)
with VLA $\lambda$=7mm observations. These observations are 
overplotted on the L$^{\prime}$ image in Figure 4.
The observed disk is quite compact - only $\sim$0.3$^{\prime\prime}$ (40 AU) in diameter.
Though this system is a close binary, we do not expect the companion to affect
the modeled parameters significantly, because our results mostly constrain the outer disk.
The compactness of the disk measured \citet{loinard2002}
is not inconsistent with our disk model since the $\lambda$=7mm observations
will only be sensitive to the densest part of the disk and where the grains are largest. 
Their measured disk mass of 0.1$M_{\sun}$ is substantially larger than our
disk model (which is uncertain as discussed previously). Note that the disk mass is strongly 
dependent on the assumed dust mass opacities; \citet{loinard2002} used a dust opacity
model based on \citet{pollack1994} which had $\kappa_{7mm}$=4.9$\times$10$^{-4}$cm$^2$g$^{-1}$ 
(P. D'Alessio private communication)). \citet{andrews2005} used $\kappa_{850\mu m}$=0.035 cm$^2$g$^{-1}$
with $\kappa$ $\propto$ $\lambda^{-1}$ which yields $\kappa_{7mm}$=0.00425 cm$^2$g$^{-1}$
when extrapolated. The difference between the $\kappa_{7mm}$ values can be attributed to
the \citet{pollack1994} dust model only considering grains up to 1$\mu$m in size while the dust model 
used by \citet{andrews2005} considered grains up to 1mm in size. Scaling the \citet{loinard2002}
 $\lambda$=7mm mass to $\kappa_{7mm}$ from \citet{andrews2005} gives $M_{disk}$=0.0095$M_{\sun}$, in
agreement with the mean disk mass of Taurus.

From the observations and modeling it is now clear that the small-scale scattered light in
L1527 is most likely due to a disk. However, the assumed parameters needed to fit the disk model
(e.g. steep density profile) may only have been necessary to enable
our simplistic treatment of disk structure to work for L1527.
Models which solve for the vertical structure of the disk self-consistently \citep[i.e.][]{dalessio1998},
given the irradiation from the central object have
a density structure that is not Gaussian (as assumed in our modeling)
but falls off less rapidly with $z$ because the temperature increases
with vertical height $z$.  Indeed, even in the context of an 
isothermal structure, the Gaussian density distribution becomes a poor 
approximation as the ratio of the vertical height to radial distance 
becomes $z/R \rightarrow 1$.  These issues deserve further investigation
with physically self-consistent radiative transfer modeling of L1527.

\subsection{Conclusions}

We have presented high-resolution L$^{\prime}$ imaging of the inner envelope of L1527
demonstrating that much can be learned of inner envelope structure and 
newly forming disks with high resolution scattered light observations of Class 0 protostars.
The observations strongly resemble the dual-cavity model constructed in
\citet{tobin2008}; however, the observations can be equally well reproduced by a 
large, highly flared disk or a refined dual-cavity structure.
However, the global scattered light structure is difficult to 
interpret as purely an outflow cavity. We therefore suggest that the bright
scattered light structure is most likely be a vertically extended disk-in-formation
with material falling onto it.

The authors wish to thank the anonymous referee for a helpful
report making the manuscript more clear, Gemini staff Astronomer R. Mason for critical
assistance in conducting the observations, B. Whitney for providing her Radiative Transfer code
to the community, and P. D'Alessio for helpful discussions. J. T. and L. H. acknowledge 
funding from HST-GO-11548.04-A and partial support from the University of Michigan.
L. L. acknowledges the support of DGAPA (UNAM), CONACyT  (Mexico),
and the Guggenheim Memorial Foundation.

{\it Facilities:} \facility{Gemini:Gillett (NIRI)}, \facility{Spitzer (IRAC)}, \facility{VLA}
\begin{small}

\end{small}

\begin{figure}
\begin{center}
\includegraphics[angle=-90, scale=0.7]{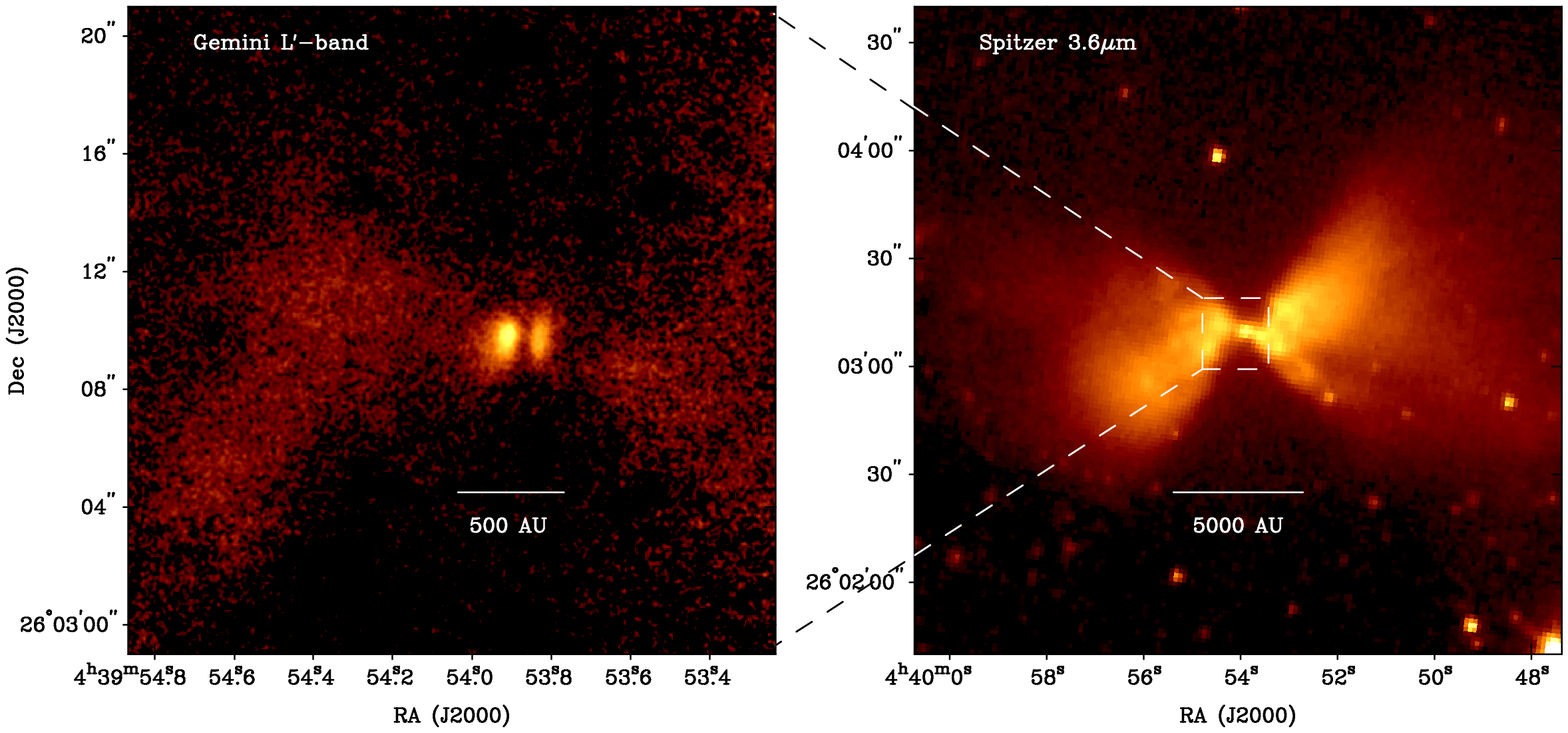}
\end{center}
\caption{Left: L$^{\prime}$ image of L1527 showing the full range of the outflow cavity viewed. Notice how the outflow
cavity rapidly widens about 6$^{\prime\prime}$ (840 AU) from the protostar. Right: Full IRAC 3.6$\mu$m image
of L1527. The region viewed in the L$^{\prime}$ observations is outlined.}
\end{figure}

\begin{figure}
\begin{center}
\includegraphics[angle=-90, scale=0.7]{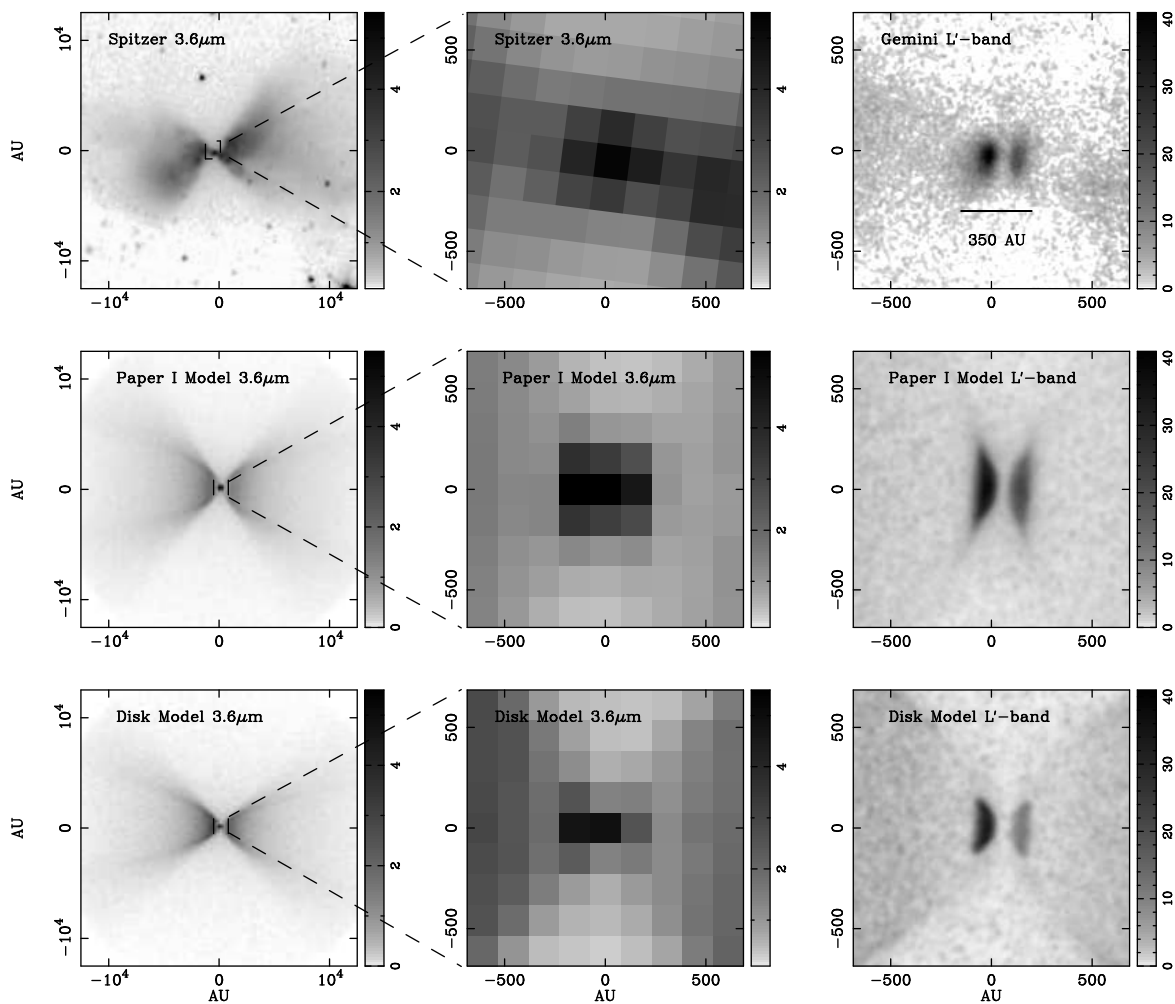}
\end{center}
\caption{Comparison of models to the observations. 
Top row: Observations of L1527 from Spitzer and Gemini.
Middle Row: Initial model of L1527 from Paper I without any knowledge of L1527 at high resolution.
Bottom Row: Disk plus single-cavity model. Note that the refined dual-cavity model images
are not shown as they are nearly identical to the disk models.}
\end{figure}

\begin{figure}
\begin{center}
\includegraphics[scale=0.5]{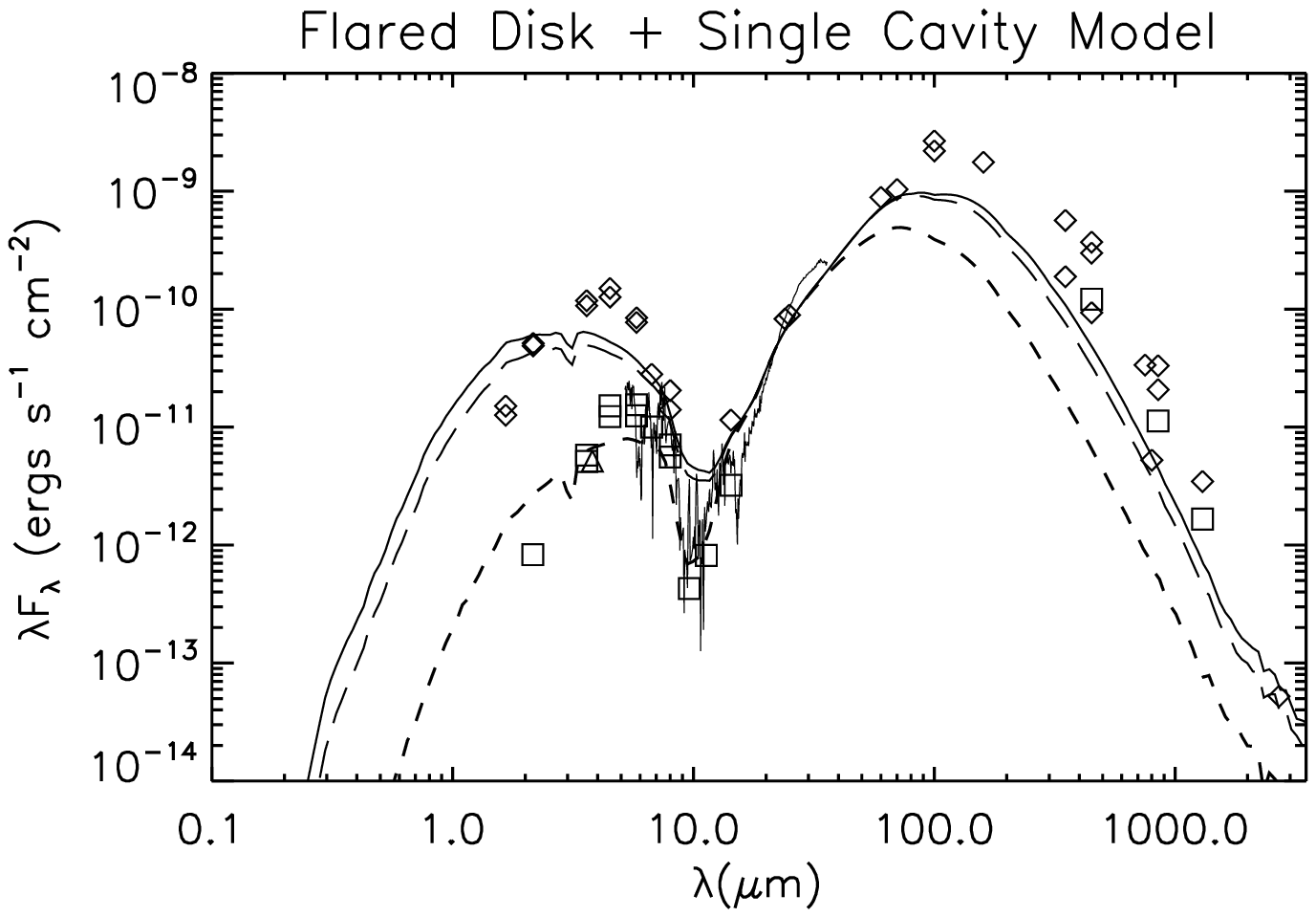}
\includegraphics[scale=0.5]{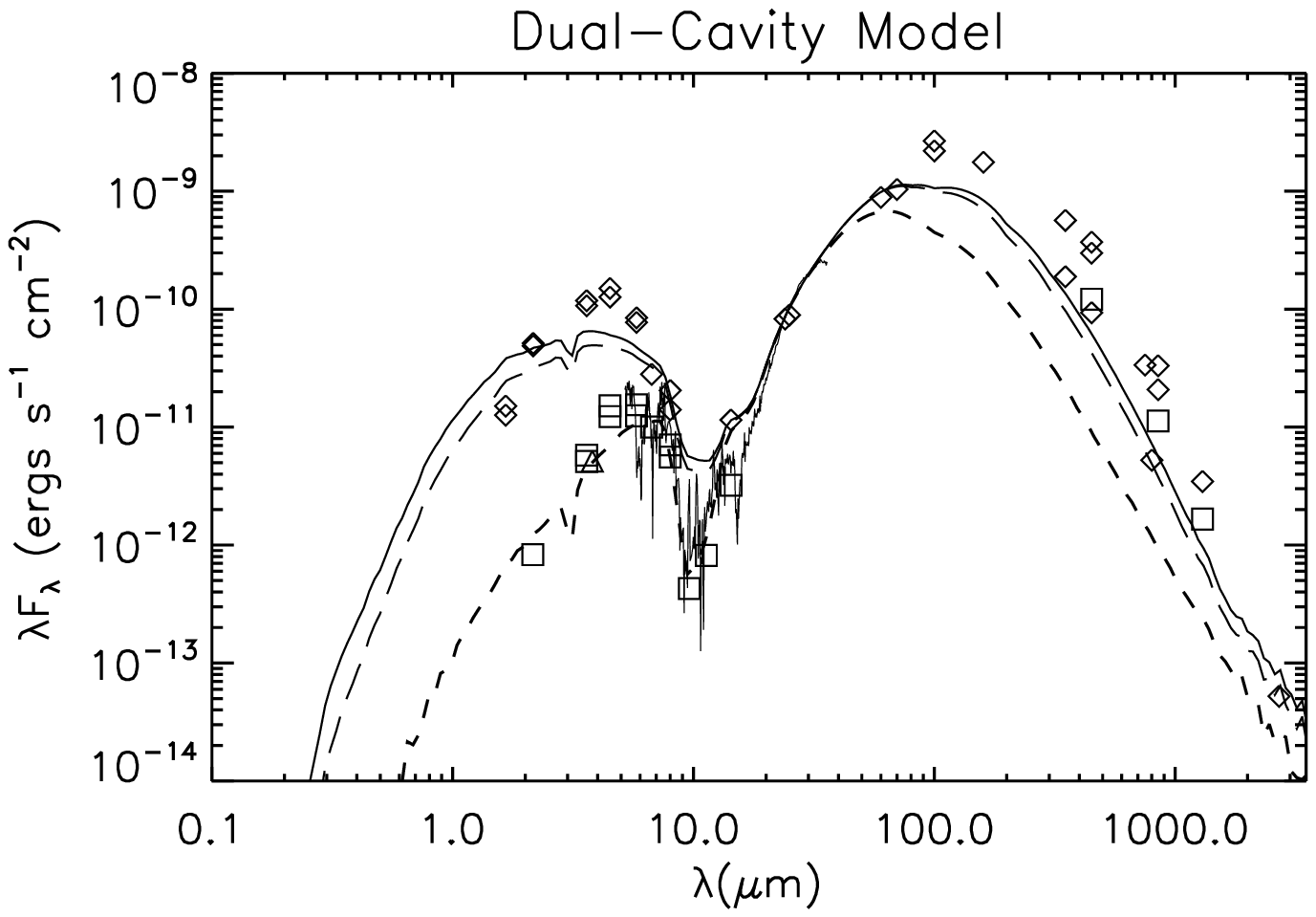}
\end{center}
\caption{
Model SEDs for L1527 with photometry and IRS spectrum from \citet[][and references therein.]{tobin2008}.
Photometry taken with apertures of 71$\farcs$4 (diamonds) and 7$\farcs$14
(boxes) (10000 AU and 1000 AU) are plotted. The triangle
at 3.8$\mu$m is the Gemini L$^\prime$ flux within 1000 AU.
The model SEDs are plotted for the Disk model (left) and the refined
dual-cavity model (right) with multiple model apertures of 10000 AU (solid line), 
6000 AU (long-dashed line), and 1000 AU (short-dashed line).  
The model is clearly deficient in flux at long wavelengths; however, external
heating is not taken into account and the dust temperatures fall below 10K in the outer
envelope emitting less in the far-IR and submm.}
\end{figure}

\begin{figure}
\begin{center}
\includegraphics{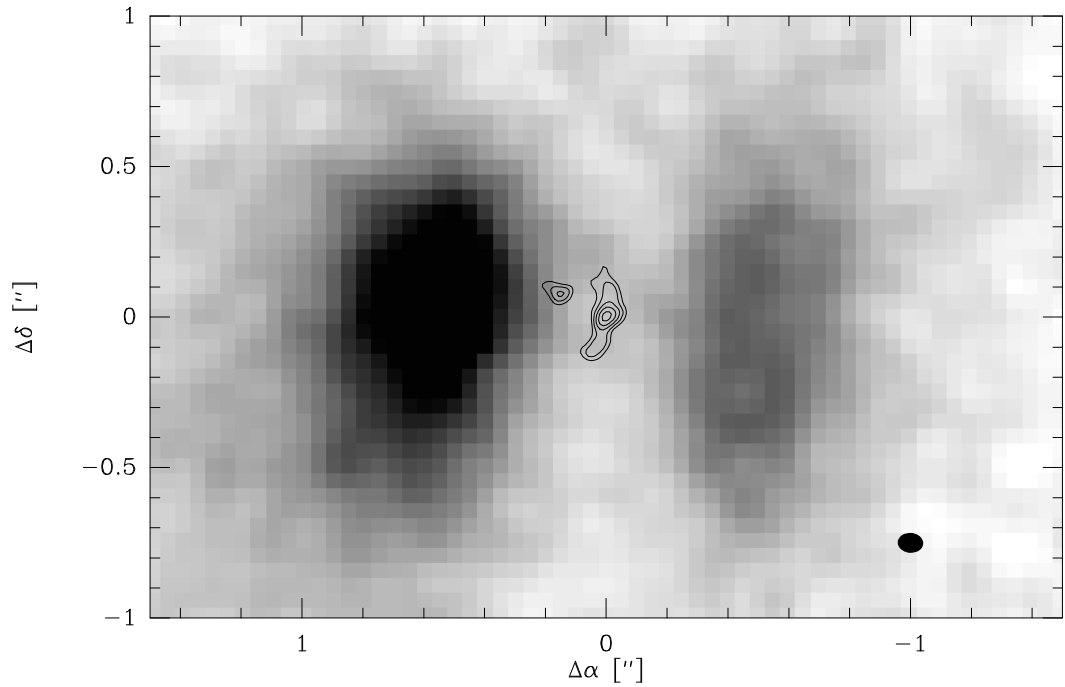}
\end{center}
\caption{L$^{\prime}$ image of L1527 with 7mm dust continuum observations (contours) from VLA observations by \citet{loinard2002}.
Notice that the position angle of the long axis of the disk is in the same direction as the disk extinction lane, $\sim$-5$^{\circ}$ East of North.
The second 7mm point source just east of the disk is the binary companion. }
\end{figure}

\begin{deluxetable}{llccl}
\tabletypesize{\scriptsize}
\tablewidth{0pt}
\tablecaption{Model parameters\label{param}}
\tablehead{   \colhead{Parameter} & \colhead{Description} & \colhead{Dual-Cavity Model} & \colhead{Disk Model} & \colhead{Parameter Use}}
\startdata
R$_{*}$(R$_{\sun}$) & Stellar radius & 2.09 & 2.09& fixed (Paper I)\\
T$_{*}$(K) & Stellar temperature & 4000 & 4000& fixed (Paper I)\\
L$_{*}$(L$_{\sun}$) & System luminosity & 2.75 & 2.75& fixed (Paper I)\\
M$_{*}$(M$_{\sun}$) & Stellar mass & 0.5 & 0.5 & fixed (Paper I)\\
M$_{disk}$(M$_{\sun}$) & Disk mass & 0.05  & 0.005 & varied (both models)\\
h(10AU) & Disk scale height at 10AU & 1.87& 1.95 & varied (both models)\\
H$_0$ & Disk scale height at R$_{*}$ & 0.0332 & 0.03& varied (both models)\\
$\alpha$ & Disk radial density exponent & 2.25  & 3.0& varied (both models)\\
$\beta$ & Disk scale height exponent & 1.25 & 1.27& varied (both models)\\
$\dot{M}_{disk}$(M$_{\sun}$ $yr^{-1}$) &  Disk accretion rate & 3.0 $\times 10^{-7}$ & 3.0 $\times 10^{-7}$ & fixed (Paper I) \\
R$_{trunc}$(R$_{*}$) & Magnetosphere co-rotation radius & 3.0 & 3.0& fixed (Paper I)\\
F$_{spot}$ & Fractional area of accretion hotspot & 0.01 & 0.01 & fixed (Paper I)\\
R$_{disk,min}$(R$_{*}$) & Disk inner radius & 14.25 & 14.25 & fixed (Paper I)\\
R$_{disk,max}$(AU) & Disk outer radius & 25  & 190  & varied (both models)\\
R$_{c}$(AU) & Centrifugal radius & 25 & 190 & varied (coupled to R$_{disk,max}$)\\
R$_{env,min}$(R$_{*}$) & Envelope inner radius & 42.75 & 42.75  & fixed (Paper I)\\
R$_{env,max}$(AU) & Envelope outer radius & 15000  & 15000 & fixed (Paper I)\\
$\dot{M}_{env}$(M$_{\sun}$ $yr^{-1}$) & Envelope mass infall rate & 1.00 $\times 10^{-5}$ & 0.8 $\times 10^{-5}$ & varied slightly (both models)\\
b$_{in}$ & Inner cavity shape exponent & 1.3 & -& varied (dual cavity model)\\
b$_{out}$ & Outer cavity shape exponent & 1.7 & 1.5 & varied (both models)\\
z$_{out}$(AU) & Outer cavity offset height & 85 & - & varied (dual cavity model)\\
$\theta_{open,in}$($^{\circ}$) & Inner cavity opening angle & 15 & - & varied (dual cavity model)\\
$\theta_{open,out}$($^{\circ}$) & Outer cavity opening angle & 15 & 20& varied (both models)\\
$\theta_{inc}$($^{\circ}$) & Inclination angle & 85 & 85 & fixed (Paper I)\\
$\rho_{c}$(g cm$^{-3}$) & Cavity density & 0 & 0& fixed (Paper I)\\
$\rho_{amb}$(g cm$^{-3}$) & Ambient density & 0 & 0& fixed (Paper I)
\enddata
\end{deluxetable}

\end{document}